\documentclass{elsart}

\usepackage[fleqn,intlimits]{amsmath}
\usepackage[dvips]{graphicx}

\newcommand{\w}{\Omega}
\newcommand{\pd}{\partial}

\newcommand{\xt}{\langle x(t) \rangle}

\begin{document}
\runauthor{G. Schmid, P. H\"anggi}
\journal{Physica A}

\begin{frontmatter}
\title{Controlling Nonlinear Stochastic Resonance by Harmonic Mixing}

\author{Gerhard Schmid\corauthref{cor1}},
\author{Peter H\"anggi}
\address{Universit\"at Augsburg, Institut f\"ur Physik,
Theoretische Physik I,
 D-86135 Augsburg, Germany}
\corauth[cor1]{Corresponding author, tel.: +49-821-598-3229,
fax:+49-821-598-3222, e-mail: Gerhard.Schmid@physik.uni-augsburg.de }

\begin{abstract}
We investigate the potential for controlling the effect of
nonlinear Stochastic Resonance (SR) by use of harmonic mixing
signals for an overdamped Brownian dynamics in a symmetric double
well potential. The periodic forcing for {\it harmonic mixing}
consists of a first signal with a basic frequency $\Omega$ and a
second, superimposed signal oscillating at twice the basic
frequency $2\Omega$. By variation of the phase difference between
these two components  and the amplitude ratios of the driving the
phenomenon of SR becomes  {\it a priori} controllable. The
harmonic mixing dynamically breaks the symmetry so that the time-
and ensemble-average assumes a non-vanishing value. Independently
of the noise level, the response can be suppressed by adjusting
the phase difference. Nonlinear SR then exhibits resonances at
higher harmonics with respect to the applied noise strength and
relative phase. The scheme of nonlinear SR via harmonic mixing can
be used to steer the nonlinear response and to sensitively measure
the internal noise strength. We further demonstrate that the full
Fokker-Planck dynamics can be well approximated by a two-state
model.
\end{abstract}
\end{frontmatter}

\section{Introduction}
\label{introduction}

{\it Stochastic Resonance} (SR) describes the phenomenon where an
incoming, generally weak signal can become amplified upon
harvesting the ambient noise in metastable, nonlinear stochastic
systems \cite{gammaitoni98}. This phenomenon is based on a
stochastic synchronization between  noise-induced hopping events
and the periodic, externally applied signal
\cite{gammaitoni98,callenbach2002a,freund2003,gammaitoni1995}. 
SR has since been observed in
an abundance of systems in physics, chemistry, engineering,
biology and biomedical sciences --- and the list of examples and
applications is still growing. In particular SR has found
widespread interest and has been  applied to many differing
applications within  biological physics \cite{hanggi2002a}. In
many situations, however, the strength of the noise acting upon a
system is not arbitrarily controllable; e.g. the strength of the
internal noise source can be so large that SR simply will not
occur, as it may happen  for SR in globally coupled ion channel
clusters of small size \cite{schmidepl,schmid2004,jungepl}.  It is therefore
of ultimate importance to devise control schemes to attain and
manipulate SR in real systems. A concept which was proposed in
prior  literature \cite{loecherprl,loecherpre} in order to enhance
or suppress the spectral power is based on a modulation of the
threshold in a discrete detector or in a bistable system
dynamics. 
This in turn results in
"breathing" oscillations of the barriers. 
By doing so,  the ``classical'' SR effect could be both
characteristically enhanced and suppressed by changing the phase
difference between the threshold modulation and the input signal.

In this work we suggest a different, although related control
scheme which we base on harmonic mixing input signal
\cite{HM,wonneberger,breymayer,marchesoni1986,goychuk1998,savelev}.
The sinusoidal input signal is superposed by a second, sinusoidal
signal with twice the frequency of the former, monochromatic input 
signal. By controlling the phase difference between these two
signal parts we obtain a powerful tool for the manipulation of SR.

\section{The model}
\label{model}

To start out, we consider the  motion of a Brownian particle in a
bistable and symmetric potential in the presence of noise and periodic
forcing. The particle is furthermore subjected to viscous friction.
With the assumption that inertia effects are negligible (overdamped
dynamics), the driven Langevin dynamics reads in scaled
units \cite{gammaitoni98,jung93}:

\begin{align}
  \label{eq:langevin}
  \frac{d}{dt}x(t)=-\frac{d}{dx}V(x) + f(t) + \xi(t)\; ,
\end{align}
with the static double-well potential given by $V(x):=\frac{1}{4}x^{4}-\frac{1}{2}x^{2}$.
The harmonic mixing driving signal $f(t)$ has the form,
\begin{align}
  \label{eq:signal}
  f(t)=A\sin( \w t) + B\sin(2 \w t + \Psi)\, ,
\end{align}
The  relative
phase difference is denoted by $\Psi$, and it is this quantity
which we shall  predominantly  use in the following as our control
parameter for steering SR. The coupling to the heat bath
is modeled by zero-mean, Gaussian white noise $\xi(t)$ with
autocorrelation function:
\begin{align}
  \label{eq:rauschen}
  \langle \xi(t) \xi(s) \rangle = 2 D \delta(t-s)\, ,
\end{align}
where $D$ denotes the noise strength.

The corresponding
Fokker-Planck equation for the probability density $P(x,t)$
\cite{hanggithomas1982,risken} is thus given by,
\begin{align}
  \label{eq:fokkerplanck}
  \frac{\pd}{\pd t}P(x,t) = \frac{\pd}{\pd x}\left[ \left(\frac{d}{dx}V(x)
  \right)
  -f(t)
%  - A \sin(\w t) - B \sin(2 \w t + \Psi)
\right] P(x,t)
  + D \frac{\pd^{2}}{\pd x^{2}} P(x,t)\, .
\end{align}

In the absence of the second signal (i.e. $B=0$) eq.
\eqref{eq:langevin} forms the archetypical model for
SR\cite{gammaitoni98}. The dependence of SR-measures such as the
spectral power amplification \cite{jung1989,jung1991a,gammaitoni1989} or the
signal-to-noise ratio \cite{mcnamarawiesenfeld89}, respectively,
exhibits a bell-shaped behavior {\it vs.} the noise strength $D$.
Moreover, due to the dynamical generalized parity symmetry
\cite{jung1989,jung1991a,gammaitoni1989,hanggi1993a} only odd higher harmonics emerge
which all exhibit the effect of SR. In contrast, for asymmetric
double-well potentials also the even numbered higher harmonics are
generated: The generation rate of the third harmonic then depicts
a characteristic noise-induced suppression
\cite{jungtalkner95,jungbartussek96}. Due to our harmonic mixing
signal,  and particularly due to the relative phase difference
$\Psi$ and the ratio of  amplitudes $A$ and $B$, we can
systematically break the symmetry dynamically and thus, control
the response at higher harmonics.

\section{Symmetry breaking in the deterministic model}
\label{deterministic}

Before we elucidate the Fokker-Planck dynamics (\ref{eq:fokkerplanck}),
the deterministic case is instructive for obtaining an understanding of the physics
of the harmonic mixing driving on the dynamics of a particle in a symmetric double-well.
The Langevin equation (\ref{eq:langevin}) thus  turns into the time-dependent deterministic equation:

\begin{align}
  \label{eq:newton}
  \frac{d}{dt}x(t) = - \frac{d}{dx}V(x) + f(t)\, .
\end{align}
In absence of any modulation, i.e. $f(t)=0$, there exist two
stable attractors at $x_{\pm}=\pm 1$ and one unstable attractor at
$x_{\rm u}=0$. For sub-threshold harmonic mixing,
$f(t)=A\sin(\Omega t) + B\sin(2 \Omega t + \Psi)$,  two
oscillatory stable orbits are formed within the potential wells.
The domains of attraction are separated by an unstable orbit,
oscillating close to the former, unstable point near the barrier.

\begin{figure}[t]
  \centering
  \includegraphics{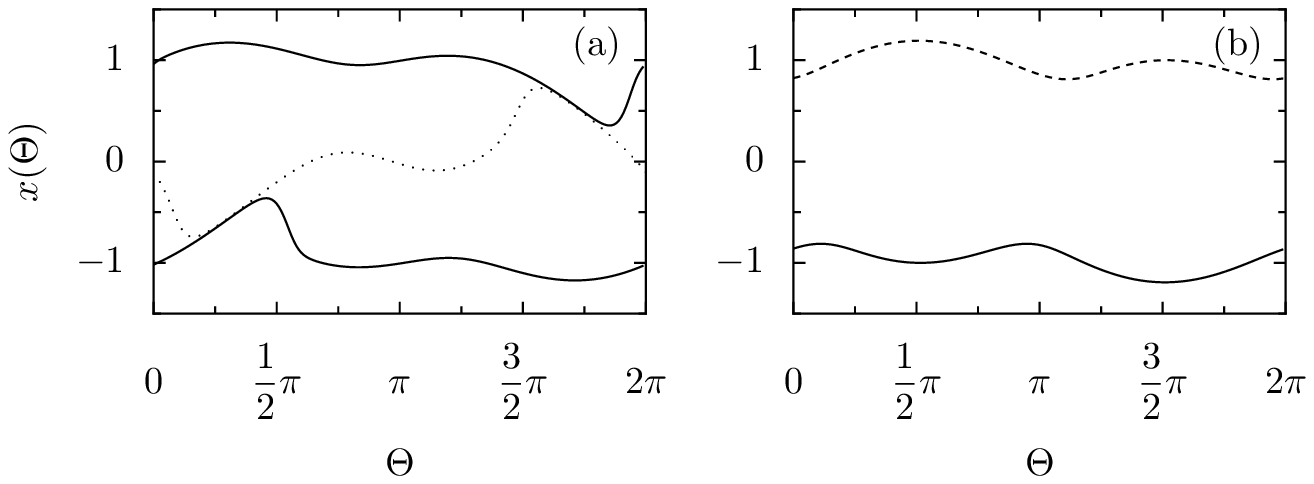}
  \caption{The stable (solid line) and unstable (dashed line) periodic orbits
  for the motion of an {\it overdamped} particle in a quartic double-well
  potential driven by a harmonic mixing signal $f(t)=0.25\ \sin(0.1\ t)
  + 0.25\ \sin(0.2\ t + \Psi)$ are plotted for $\Psi=0$ in Fig.~(a)
  with $\Theta:= 0.1\ t (\mod 2\pi)$.
  In contrast, there exists also a parameter regime for the  phase difference $\Psi$ where
  only one stable orbit exists (depicted in panel (b)). For example, the orbit is
  located in the left potential well for $\Psi=\pi/2$ (solid line) and
  in the right well for $\Psi=3\pi/2$ (dashed line); harmonic
  mixing thus causes symmetry breaking. }
  \label{fig:deterministic}
\end{figure}

Upon increasing the amplitudes $A$ and $B$, the oscillations of
the stable and instable orbits become larger; consequently, the
corresponding orbits approach each other. At even stronger driving
the situation changes drastically and the variation of the phase
difference $\Psi$ possesses  salient effects: the symmetry
breaking by the harmonic mixing signal becomes evident: In
Fig.~\ref{fig:deterministic} the amplitudes $A$ and $B$ both equal
the barrier height $1/4$ and the basic frequency is $\w=0.1$. For
phases $\Psi$ around $0$ or $\pi$ two stable and one unstable
orbits are present, whereas for $\Psi=\pi/2$ or $\Psi=3\pi/2$
there is only one attractor located in either the left or in the
right potential well of the static potential, respectively. The
reason for this symmetry breaking is the interplay between the
asymmetry of the harmonic mixing signal and the non-linearity of
the quartic double-well potential.

For large signal amplitudes and arbitrary phase differences, there
occurs only one stable periodic orbit which spreads over both potential wells
(not depicted).

\section{Two-State model}
\label{twostate}

In view of  our findings for the deterministic dynamics, we  expect,
that  the noisy system exhibits SR similarly to that occurring in asymmetric
potentials driven by sinusoidal signals \cite{gammaitoni98,jung93}. In
order to check the former statement we have numerically solved the
continuous Fokker-Planck model and developed an approximate treatment
for \eqref{eq:langevin} in terms of  a two-state model.

For small driving frequencies, i.e. for frequencies which are much
smaller than the noise induced hopping rate, the adiabatic potential modulation can be
invoked. Applying Kramers rate formula
\cite{kramers40,hanggi1985,hanggi1986,hanggiRMP90} for the transition rates among
potential wells we find to leading
order in the driving amplitudes \cite{jung89} the results:
\begin{align}
  \label{eq:rates}
  k_{\pm}(t)=k_{0} \  \exp \left\{ \pm \frac{A}{D} \sin (\w t) \pm \frac{B}{D}
  \sin(2 \w t + \Psi) \right\}\, ,
\end{align}
wherein $k_{0}$ is the Kramers rate of the unperturbed symmetric
system, i.e.  $k_{0} := 1 / (\pi \sqrt{2}) \exp \left\{ -1 / (4 D)
\right\}$. The occupation probabilities $p_{\pm}(t)$ for the two
states $x_{\pm}=\pm 1$ obey the following master equation
\cite{mcnamarawiesenfeld89,casado2003}:
\begin{align}
  \label{eq:masterequation}
  \frac{d}{dt} p_{\pm}(t) = k_{\pm}(t) p_{\mp}(t) - k_{\mp}(t) p_{\pm}(t)\, .
\end{align}
Due to normalization of probabilities, i.e.
$p_{+}(t)+p_{-}(t)=1$, the differential equation for the
mean value ($\xt = p_{+} - p_{-}$) reads
\begin{align}
  \label{eq:differentialequation}
  \frac{d}{dt} \xt = - \left[ k_{+}(t) + k_{-}(t) \right] \xt + k_{+}(t) -
  k_{-}(t)\, .
\end{align}
Assuming $A\approx B$ and $A/D\ll 1$ the asymptotic, periodic long
time solution of eq.~\eqref{eq:differentialequation} can be
expanded beyond linear response into a series with respect to the
ratios $A/D$ and $B/D$. Next, in order to identify higher
harmonics, we expand $\xt$ into a Fourier series:
\begin{align}
  \label{eq:fourier}
  \xt = \gamma_{0} + \sum_{n=1}^{\infty}
  \gamma_{n} \sin( n \w t + \phi_{n})\,  ,
\end{align}
with corresponding  Fourier coefficients ${\gamma_{n}}$ and phase
lags ${\phi_{n}}$. The spectral amplification factors $\eta_{n}$,
which are defined as ratio of the output power stored at the
corresponding higher harmonic driving frequency
  to the input power, are given by
\begin{align}
  \label{eq:amplification}
  \eta_{n}=\frac{\gamma_{n}^{2}}{A^{2}+B^{2}}, \ n=1,2,... \ .
\end{align}

\subsection{The time- and ensemble-averaged mean value}
\begin{figure}[t]
  \centering
  \includegraphics{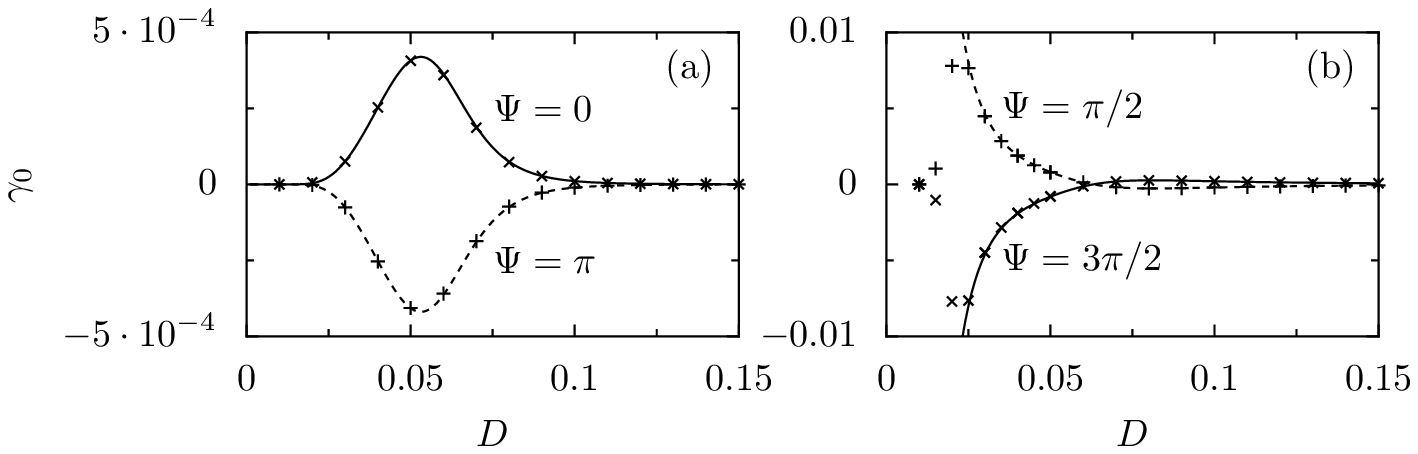}
  \caption{Time- and ensemble-averaged mean value $\gamma_{0}$ of the
  nonlinear response of the two-state system driven by a harmonic mixing signal
    with amplitudes $A=B=0.01$, fundamental frequency $\w=0.01$ and
    phase differences $\Psi=0, \pi/2,
    \pi$ and $3\pi/2$.
    The lines correspond to the analytic solution, i.e. eq.~\eqref{eq:gamma0},
    while the two symbols ("$+$" and "$\times$" ) belong to the corresponding
    numerical solution of eq.~\eqref{eq:differentialequation}.
    This driving induced zero-frequency response $\gamma_{0}$ exhibits versus noise strength $D$
    a bell-shaped behavior, similar to the behavior of  Stochastic Resonance.
    Interestingly, for specific noise levels and chosen relative phases $\Psi$ the
    symmetry can be restored, cf. panel 2(b) and Fig.~\ref{fig:contour}(a).
  }
  \label{fig:meanvalue}
\end{figure}

We start our discussion with the zeroth order Fourier coefficient $\gamma_{0}$, namely the time- and noise averaged
mean value. This nonlinear response reads in leading order:
\begin{align}
  \label{eq:gamma0}
  \gamma_{0} = &\ \frac{A^{2} B}{D^{3}}\ \frac{1}{8} \
  \frac{1}{(4 k_{0}^{2} +
    \Omega^{2} ) (k_{0}^{2} + \Omega^{2})} \times  \nonumber \\
  & \bigg\{  3 k_{0} \Omega^{3} \cos \Psi
  + ( 8 k_{0}^{4} + 4 k_{0}^{2} \Omega^{2} - \Omega^{4}) \sin \Psi \bigg\}\, .
\end{align}
Please note that generally $\gamma_{0}$ differs from zero. This is
so, because the unbiased, but   asymmetric input signal,
possessing particularly nonvanishing time-averaged odd numbered
higher moments $ n \geq 3$ dynamically breaks the symmetry of the
system
\cite{loecherprl,loecherpre,mahatojayannavar,mahatojayannavar1998,hanggi1996a,borromeo,golding00}.
For illustration, we depict this driving induced, nonvanishing
mean $\gamma_{0}$ for $f(t)=0.01 \sin(0.01 t) + 0.01 \sin(0.02 t +
\Psi)$ and different relative phases $\Psi$ in
Fig.~\ref{fig:meanvalue}. For a phase difference $\Psi=0$, the
accumulation in the state "$+$" increases initially, reaches a
maximum, and then decreases as the noise strength $D$ is increased
further, cf. Fig. \ref{fig:meanvalue}(a). At an optimum noise
level $D$ the accumulation in one state is extremal. Similar to
the phenomenon of Stochastic Resonance, this effect manifests
itself by a synchronization of noise-activated hopping events
between the two metastable states and the driving force $f(t)$.

Upon changing the relative phase difference $\Psi$ we thus can
control the dynamical asymmetry of the harmonic mixing driving
signal and therefore the time averaged mean value $\gamma_0$. As a
consequence for $\Psi=0$ and $\pi$ the accumulation in the states
"$+$ and $-$" undergo an SR - like behavior. For other phase
differences $\Psi$, the time averaged mean value vanishes at
certain noise strengths; thus, the symmetry in the system can,
dynamically, be restored accidentally at selected parameter
choices, cf. Fig.~\ref{fig:meanvalue}(b). According to the
expansion in $A/D$ and $B/D$, the analytic solution worsens for
small noise strengths $D$, this feature is  apparent in
Fig.~\ref{fig:meanvalue}(b).

\subsection{Spectral amplification factors}

\begin{figure}[t]
  \centering
  \includegraphics{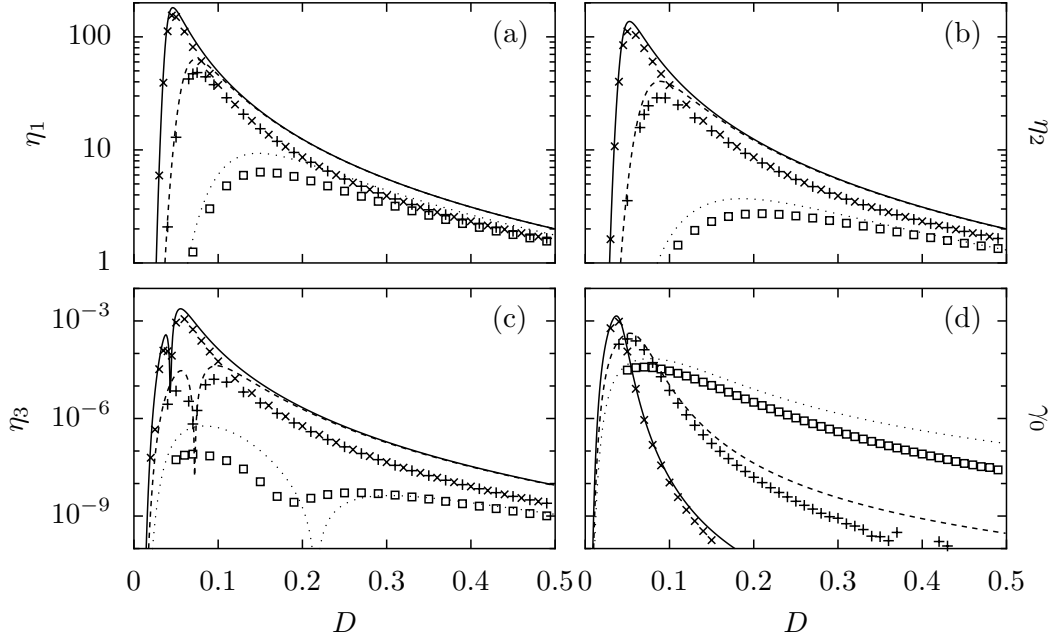}
  \caption{The dependence of the spectral power amplification factors (a)-(c) and the
  time averaged mean value (d) versus the noise strength $D$ is depicted
  for the driving amplitudes $A=B=0.01$ at vanishing relative phase $\Psi=0$ and at the fundamental
  driving frequency  $\Omega=0.001$:
  analytic estimate (solid line), corresponding numerical Fokker-Planck solution (crosses "$\times$"). The same for
  the driving fundamental at $\Omega=0.01$: analytic estimate (dashed line), numerical solution ("+" signs).
  Likewise, the same for  the high frequency drive at $\Omega=0.1$: analytic estimate
    (dotted line), numerical Fokker-Planck solution (squares). Note that at large  driving frequencies there is a good
    agreement between analytic results (lines) and the numerical
    results (symbols) for the
    Fokker-Planck equation \eqref{eq:fokkerplanck}.}
  \label{fig:amplification}
\end{figure}

The spectral amplification factors
(\ref{eq:amplification}) at the first and second harmonic of the
system output
are evaluated to leading, non-vanishing order as:
\begin{align}
  \label{eq:eta1}
  \eta_{1} =\frac{A^{2}}{D^{2}} \ \frac{1}{A^{2}+B^{2}} \
  \frac{4k_{0}^{2}}{4k_{0}^{2} + \Omega^{2}}\, , \quad
  \eta_{2} =\frac{B^{2}}{D^{2}} \ \frac{1}{A^{2}+B^{2}} \
  \frac{k_{0}^{2}}{k_{0}^{2} + \Omega^{2}}.
\end{align}
We observe that, within this two-state approximation scheme,
$\gamma_{1}$ depends in lowest order only on $A/D$ (linear response limit).
Likewise, the spectral amplification at $2\w$ is determined in linear response by the second harmonic
 component of the harmonic mixing signal, yielding the spectral
amplification of the second harmonic $\eta_{2}$. The two
components of the driving do not interact with each other in this lowest
order, particularly because of the suppression of  even-numbered higher harmonic
generation in symmetric systems driven by sinusoidal signals. Therefore, SR
manifests itself at both frequencies with the well-known bell-shaped amplification behavior, cf.
Fig.~\ref{fig:amplification}(a) and (b).

For the generation of the third higher harmonic, however, the two parts of the
harmonic mixing signal do interact, and, in lowest, leading order, $\eta_{3}$ is given by
the expression:
{\allowdisplaybreaks
  \begin{align}
    \label{eq:eta3}
    \eta_{3} =&\ \frac{1}{D^{6}} \ \frac{1}{A^{2}+B^{2}} \ \Big
    [ \big(k_{0}^{2}+ \Omega^{2}\big) \big(4 \ k_{0}^{2}+ \Omega^{2}\big)^{2} \big(4 \
    k_{0}^{2}+ 9 \ \Omega^{2}\big) \Big]^{-1}   \times \nonumber \\[0.05cm]
    & \Biggl\{   A^{6}\ \frac{1}{144}\ k_{0}^{2}  \ \big(\Omega^{2}+16 \
    k_{0}^{2}\big) \big(k_{0}^{2}+ \Omega^{2}\big) \big(4 \ k_{0}^{2}+ \Omega^{2}\big)
    \nonumber   \\
    & - A^{4}B^{2} \ k_{0}^{2} \ \Bigg[ \frac{1}{12} \Big( \Omega^{6} + 64 \ k_{0}^{6} +
    36 \ k_{0}^{4} \Omega^{2} - 9 \  k_{0}^{2} \Omega^{4} \Big) \cos^{2} \Psi \nonumber\\
    & \hspace{2cm} + \frac{1}{2} \ k_{0}\ \w^{3} \ \Big( \Omega^{2} - 2 \
    k_{0}^{2}\Big) \  \sin \Psi \ \cos \Psi \nonumber \\
    & \hspace{2cm} - \frac{1}{24} \ \Big( 64 \ k_{0}^{6} +36 \ k_{0}^{4} \Omega^{2} -
    9 \ k_{0}^{2} \Omega^{4} + \Omega^{6} \Big) \Bigg] \nonumber   \\
    & + A^{2}B^{4} \ \frac{1}{16} \
    k_{0}^{2} \ \big( \Omega^{4}-7\ k_{0}^{2}\ \Omega^{2}+16\ k_{0}^{4}
    \big)  \big(4 \ k_{0}^{2}+ \Omega^{2}\big)  \Biggr\}\, .
  \end{align}
}

Just as for the case with an asymmetric double well potential
\cite{jungbartussek96,bartussek1994a,marchesoni1999}, the spectral
amplification at the third harmonic exhibits in our case a
noise-induced suppression. This characteristic suppression at a
tailored noise strength
 depends on the driving frequency $\Omega$ and is accompanied with a corresponding
$\pi$-phase jump (not depicted). In
Fig.~\ref{fig:amplification}(c) we depict this behavior for
amplitudes $A=B=0.01$, and a vanishing relative phase difference
$\Psi=0$ and for different fundamental frequencies.
Agreement with
the  two-state theory is best at moderate fundamental driving
frequencies; 
this  corroborates with the fact that the linear
response analysis and its corrections to higher orders indeed work
best at moderate-to-large frequencies  and increasingly fails at very small
frequencies \cite{casadoLRT,casadoLRT2}.

\subsection{Comparison with the Fokker-Planck treatment}

Additionally, we have numerically integrated the Fokker-Planck equation \eqref{eq:fokkerplanck}
and evaluated the time-periodic, asymptotic mean value $\xt$ together with an expansion  according to
eq.~\eqref{eq:fourier} into a Fourier series.
The results are depicted in Fig.~\ref{fig:amplification}.
There is  good agreement between the analytic
solution of the two-state approximation and the numerical solution of the
continuous-state problem. Although the two-state approximation, i.e. the
Kramers-rate approximation fails for large driving frequencies and
large noise strengths, respectively,  there is nevertheless still  qualitative good
agreement, cf. Fig.~\ref{fig:amplification}.

\section{Controlling nonlinear SR with noise and relative phase $\Psi$}
\label{controlling}

\begin{figure}[t]
  \centering
  \includegraphics{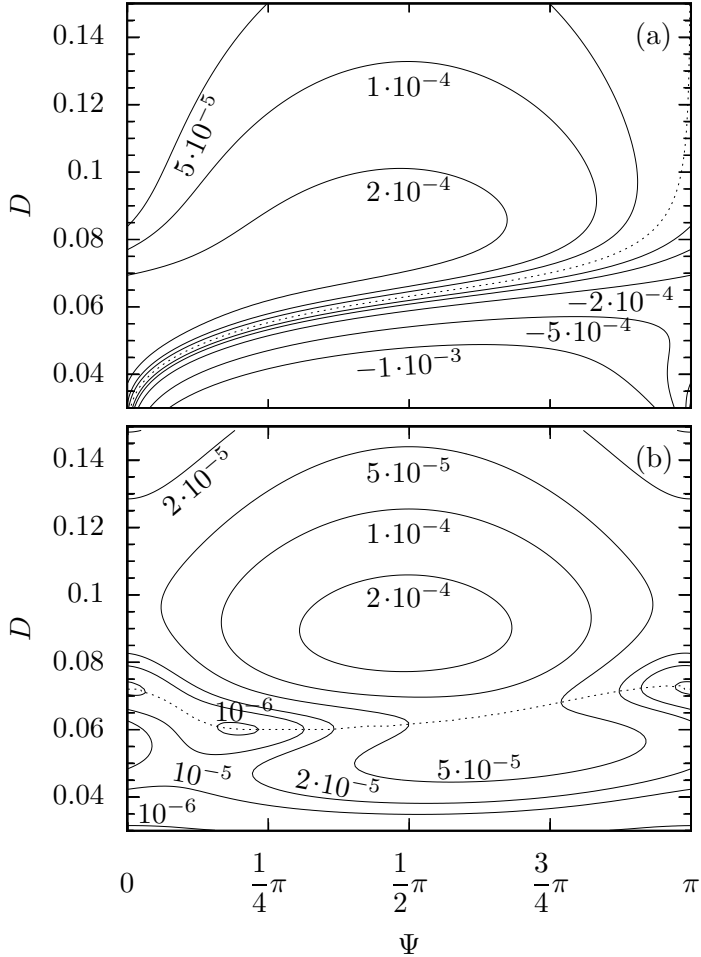}
  \caption{The contour plot of the time averaged mean value
  $\gamma_{0}$ (a) and of the spectral
    amplification factor of the third harmonic $\eta_{3}$ (b) are
    depicted for varying phase difference $\Psi$ and noise strength
    $D$ according eq.~\eqref{eq:gamma0} and eq.~\eqref{eq:eta3} ($A=B=0.01,\, \w=0.01$).
    The two dashed lines indicate the zero contour-line, meaning the
    symmetry restoring condition in (a) and the corresponding line in panel (b)
    the regime of noise-induced suppression of
    $\eta_{3}$.}
  \label{fig:contour}
 \end{figure}

Within the range of small harmonic mixing driving amplitudes, where the agreement of the
two-state and the continuous system is very good,
the time-averaged and noise averaged mean value
$\gamma_{0}$ and the spectral amplification factor of the third harmonic $\eta_{3}$
depict a striking dependency on the relative phase $\Psi$; in contrast
 the amplification factors of
the first and second harmonic generations are in lowest order
independent on the phase difference. This is because these former
quantifiers depend nonlinearly on the driving amplitudes
(nonlinear response regime). In Fig.~\ref{fig:contour} this
dependence of the time averaged mean value $\gamma_{0}$ (a) and
the third spectral amplification factor $\eta_{3}$ (b) are plotted
versus the noise strength and the relative phase difference by
means of contour-line plots. Because a shift of $\pi$ will not
change the spectral amplification factors and only inverts the
sign of the time averaged mean value $\gamma_{0}$, it is
sufficient to vary $\Psi$ in the range from $0$ to $\pi$.

As noted above, the mean value vanishes for certain, tailored
noise strengths $D$ and relative phases $\Psi$, cf.
Fig.~\ref{fig:meanvalue}(b). The resulting zero-lines converge for
large noise strengths to multiples of $\pi$, cf.
Fig.~\ref{fig:contour}(a). Interestingly enough, for every phase
difference there exists  only one value of noise strength for
which $\gamma_{0}$ vanishes and, thus, symmetry restoring occurs
accidentally. This feature can be used to determine and
characterize  sensitively the operating internal noise level in
metastable systems. Additionally, by changing the relative phase
difference the time averaged mean value and, consequently, the
output power of the dynamically induced bias value of the response
signal can be controlled. A maximum enhancement of $\gamma_{0}$ is
obtained for relative  phases $\Psi$ around $\pi/2$ and $3\pi/2$,
respectively.

By variation of the phase difference $\Psi$, the noise strength $D$ at which
suppression takes place could be controlled as well,
cf. Fig. \ref{fig:contour}(b). Yet another feature to be obtained upon controlling the
relative phase difference $\Psi$ is a large enhancement of $\eta_{3}$ up to a factor of ten.

%%%%%%%%%%%%%%%%%%%%%%%%%%%%%%%%%%%%%%%%%%%%%%%%%%%%%%%%%%%%%%%%%%%%%%%%%%%%%%%
% Summary %%%%%%%%%%%%%%%%%%%%%%%%%%%%%%%%%%%%%%%%%%%%%%%%%%%%%%%%%%%%%%%%%%%%%
%%%%%%%%%%%%%%%%%%%%%%%%%%%%%%%%%%%%%%%%%%%%%%%%%%%%%%%%%%%%%%%%%%%%%%%%%%%%%%%

\section{Summary}

We have investigated the influence of a harmonic mixing signal
on the phenomenon of nonlinear
Stochastic Resonance \cite{gammaitoni98,casado2003} for a Brownian dynamics in a double well.
In the deterministic limit of harmonic mixing driving we can distinguish three
situations: for small driving amplitudes the particle oscillates in one of the wells,
depending on the initial starting value.
In the range of large amplitudes the
oscillation extend over both wells.
For moderate driving amplitudes, however,  a
symmetry breaking occurs: Independent on the initial starting values the motion dwells
only one specific well. By varying the relative phase difference between the two components of the
mixing signal we can selectively control the dynamics in one of the two wells.

For the phenomenon of nonlinear Stochastic Resonance we monitor
the nonlinear response due to harmonic mixing versus the noise
strength $D$. Despite the somewhat coarse nature of the applied
two-state approximation, it nevertheless provides very good
agreement for the dynamics of the full Fokker-Planck dynamics; it
is only for very small frequencies and/or large noise strength
where the approximation starts to fail. The analytic estimate
predicts a dynamical symmetry breaking which can be selectively
controlled by the relative phase between the two driving modes and
the noise strength $D$.

The spectral amplification measures of the higher harmonics exhibit the
characteristic features of nonlinear SR in systems possessing an asymmetry.
At selected noise strengths
and relative phase differences the time averaged mean value accidentally vanishes thereby restoring
the symmetry via the combined action of noise and driving. The dynamically induced
bias value and the spectral amplification factor of the third
harmonic generation depend sensitively on the relative phase difference of the two
sinusoidal input signals. This can be used from a technological viewpoint to selectively control
the enhancement and the suppression, respectively, of the nonlinear
system response up to factor of ten.
Moreover, the dynamically induced restoration of symmetry can be harvested to measure very sensitively
the internal noise strength in a symmetric system.

%%%%%%%%%%%%%%%%%%%%%%%%%%%%%%%%%%%%%%%%%%%%%%%%%%%%%%%%%%%%%%%%%%%%%%%%%%%%%%%
% Acknowledgement %%%%%%%%%%%%%%%%%%%%%%%%%%%%%%%%%%%%%%%%%%%%%%%%%%%%%%%%%%%%%
%%%%%%%%%%%%%%%%%%%%%%%%%%%%%%%%%%%%%%%%%%%%%%%%%%%%%%%%%%%%%%%%%%%%%%%%%%%%%%%

\section{Acknowledgement}
\label{acknowledgement}

The authors gratefully acknowledge the support of this work by the Deutsche
Forschungsgemeinschaft, SFB 486, project A-10.

% % Please send figures with disk, or separately if
% % if it is an e-mail submission. (Good photo or India ink drawing.)


\begin{thebibliography}{999}
\bibitem{gammaitoni98}L. Gammaitoni, P. H\"anggi, P. Jung, F. Marchesoni,
  Rev. Mod. Phys. 70 (1998) 223.

\bibitem{callenbach2002a}
  L. Callenbach, P. H\"anggi, S. Linz, J.~A. Freund, L. Schimansky-Geier,
  Phys. Rev.  E 65 (2002) 051110.
  
\bibitem{freund2003}
J. A. Freund, L. Schimansky-Geier, P. H\"anggi, Chaos 13 (2003)
225.

\bibitem{gammaitoni1995}
L. Gammaitoni, F. Marchesoni, S. Santucci, Phys. Rev. Lett. 74
(1995) 1052.

\bibitem{hanggi2002a}
P. H\"anggi, ChemPhysChem  3 (2002) 285.

\bibitem{schmidepl} G. Schmid, I. Goychuk, P. H\"anggi,
  Europhys. Lett. 56 (2001) 22.

\bibitem{schmid2004}
G. Schmid, I. Goychuk, P. H\"anggi, S. Zeng, P. Jung, Fluct. Noise Lett. 4 (2004) L33.

\bibitem{jungepl} P. Jung, J. Shuai, Europhys. Lett. 56 (2001) 29.


\bibitem{loecherprl} L. Gammaitoni, M. L\"ocher, A. R. Bulsara,
  P. H\"anggi, J. Neff, K. Wiesenfeld, W. L. Ditto, M. E. Inchiosa,
  Phys. Rev. Lett. 82 (1999) 4574.

\bibitem{loecherpre} M. L\"ocher, M. E. Inchiosa, A. R. Bulsara,
  K. Wiesenfeld, L. Gammaitoni, P. H\"anggi, W. L. Ditto, Phys. Rev. E 62 (2000) 317.

\bibitem{HM}
K. Seeger, W. Maurer, Solid State Commun.  27 (1979) 603.

\bibitem{wonneberger}
W. Wonneberger, H.~J. Breymayer, Z. Phys. B 56 (1984) 241.

\bibitem{breymayer}
H.~J. Breymayer, H. Risken, H.~D. Vollmer, W. Wonneberger,
Appl. Phys. B 28 (1982) 335.

\bibitem{marchesoni1986}
F. Marchesoni, Phys. Lett. 119 (1986) 221.

\bibitem{goychuk1998}
I. Goychuk, P. H\"anggi, Europhys. Lett. 43 (1998) 503.

\bibitem{savelev}
S. Savel'ev, F. Marchesoni, P. H\"anggi, F. Nori, Europhys. Lett. 67 (2004) 179.

\bibitem{jung93} P. Jung, Phys. Rep. 234 (1993) 175.

\bibitem{hanggithomas1982}
P. H\"anggi, H. Thomas, Phys. Rep. 88 (1982)  207.

\bibitem{risken}
H. Risken, The Fokker-Planck-Equation, Springer Series
in Synergetics, Vol. 18, Springer Berlin, New York, 1984.

\bibitem{jung1989}
P. Jung, P. H\"anggi, Europhys. Lett. 8 (1989) 505.

\bibitem{jung1991a}
P. Jung, P. H\"anggi, Phys. Rev. A 44 (1991) 8032.

\bibitem{gammaitoni1989}
  L. Gammaitoni, F. Marchesoni, E. Menichella-Saetta, S. Santucci,
  Phys. Rev. Lett. 62 (1989) 349.

\bibitem{mcnamarawiesenfeld89}
  B. McNamara, K. Wiesenfeld, Phys. Rev. A
  39 (1989) 4854.


\bibitem{hanggi1993a}
P. H\"anggi, P. Jung, C. Zerbe, F. Moss, J. Stat. Phys. 70 (1993) 25.

\bibitem{jungtalkner95}
P. Jung, P. Talkner, Phys. Rev. E 51 (1995)
  2640.

\bibitem{jungbartussek96}
P. Jung, R. Bartussek, in Fluctuations and Order: The New
    Synthesis, edited by M. Millonas, Springer, New York, (1996) 35.

\bibitem{kramers40} H. A. Kramers, Physica 7 (1940) 284.

\bibitem{hanggi1985} 
P. H\"anggi, J. Stat. Phys. 42 (1986) 105.

\bibitem{hanggi1986}
P. H\"anggi, J. Stat. Phys. 44 (1986) 1003.

\bibitem{hanggiRMP90}
P. H\"anggi, P. Talkner, M. Borkovec,
  Rev. Mod. Phys.  62 (1990) 251.

\bibitem{jung89} P. Jung, Z. Phys. B 76 (1989) 521.

\bibitem{casado2003} J. Casado-Pascual, J. G\'omez-Ord\'o\~{n}ez,
  M. Morillo, P. H\"anggi, Phys. Rev. Lett. 91 (2003) 210601.

\bibitem{mahatojayannavar} M. C. Mahato, A. M. Jayannavar, Phys. Rev E
    55 (1997) 3716.

\bibitem{mahatojayannavar1998}
M. C. Mahato, A. M. Jayannavar, Physica A
    248 (1998) 138.

\bibitem{hanggi1996a}
P. H\"anggi, R. Bartussek, P. Talkner, J. {\L}uczka, Europhys. Lett.  35 (1996) 315.

\bibitem{borromeo}
 M. Borromeo, F. Marchesoni, Physica A (this issue).


\bibitem{golding00} B. Golding, L. I. McCann, M. I. Dykman, in
    Stochastic and Chaotic Dynamics in the Lakes: STOCHAOS, edited by
    D. S. Broomhead, E. A. Luchinskaya, P. V. E. McClintock, T. Mullin, 34
    (2000).

 \bibitem{bartussek1994a}
 R. Bartussek, P. H\"anggi, P. Jung, Phys. Rev. E 49 (1994) 3930.

\bibitem{marchesoni1999} F. Marchesoni, F. Apostolico, S. Santucci,
  Phys. Rev. E 59 (1999) 3958.

\bibitem{casadoLRT}
J. Casado-Pascual, J. G\'{o}mez-Ordo\~{n}ez, M. Morillo, P. H\"anggi,
Europhys. Lett. 58 (2002) 342.

\bibitem{casadoLRT2}
J. Casado-Pascual, C. Denk, J. G\'{o}mez-Ordo\~{n}ez, M. Morillo, P. H\"anggi,
Phys. Rev. E 67 (2003) 036109.

\end{thebibliography}
\end{document}